\RequirePackage{lineno}
\documentclass[floatfix,twocolumn,preprintnumbers,amsmath,amssymb,prl,superscriptaddress]{revtex4} 
\usepackage{url}
\usepackage{graphicx}
\usepackage{dcolumn}
\usepackage{xspace}
\usepackage{bm}
\usepackage{soul}
\usepackage{ulem}
\usepackage{xcolor}

\newcommand{\pt}{\mbox{$p_{\rm{T}}$}\xspace}
\newcommand{\snn}{\mbox{$\sqrt{s_{\rm NN}}$}\xspace}
\newcommand{\hyt}{\mbox{${}^3_\Lambda \rm{H}$}\xspace}
\newcommand{\hyh}{\mbox{${}^4_\Lambda \rm{H}$}\xspace}
\usepackage{color}
\usepackage{natbib}
\begin{document}
\raggedbottom
\title{
Measurements of ${}^3_\Lambda \rm{H}$ and ${}^4_\Lambda \rm{H}$ Lifetimes and Yields in Au+Au Collisions in the High Baryon Density Region
}

\affiliation{Abilene Christian University, Abilene, Texas   79699}
\affiliation{AGH University of Science and Technology, FPACS, Cracow 30-059, Poland}
\affiliation{Alikhanov Institute for Theoretical and Experimental Physics NRC "Kurchatov Institute", Moscow 117218}
\affiliation{Argonne National Laboratory, Argonne, Illinois 60439}
\affiliation{American University of Cairo, New Cairo 11835, New Cairo, Egypt}
\affiliation{Brookhaven National Laboratory, Upton, New York 11973}
\affiliation{University of California, Berkeley, California 94720}
\affiliation{University of California, Davis, California 95616}
\affiliation{University of California, Los Angeles, California 90095}
\affiliation{University of California, Riverside, California 92521}
\affiliation{Central China Normal University, Wuhan, Hubei 430079 }
\affiliation{University of Illinois at Chicago, Chicago, Illinois 60607}
\affiliation{Creighton University, Omaha, Nebraska 68178}
\affiliation{Czech Technical University in Prague, FNSPE, Prague 115 19, Czech Republic}
\affiliation{Technische Universit\"at Darmstadt, Darmstadt 64289, Germany}
\affiliation{ELTE E\"otv\"os Lor\'and University, Budapest, Hungary H-1117}
\affiliation{Frankfurt Institute for Advanced Studies FIAS, Frankfurt 60438, Germany}
\affiliation{Fudan University, Shanghai, 200433 }
\affiliation{University of Heidelberg, Heidelberg 69120, Germany }
\affiliation{University of Houston, Houston, Texas 77204}
\affiliation{Huzhou University, Huzhou, Zhejiang  313000}
\affiliation{Indian Institute of Science Education and Research (IISER), Berhampur 760010 , India}
\affiliation{Indian Institute of Science Education and Research (IISER) Tirupati, Tirupati 517507, India}
\affiliation{Indian Institute Technology, Patna, Bihar 801106, India}
\affiliation{Indiana University, Bloomington, Indiana 47408}
\affiliation{Institute of Modern Physics, Chinese Academy of Sciences, Lanzhou, Gansu 730000 }
\affiliation{University of Jammu, Jammu 180001, India}
\affiliation{Joint Institute for Nuclear Research, Dubna 141 980}
\affiliation{Kent State University, Kent, Ohio 44242}
\affiliation{University of Kentucky, Lexington, Kentucky 40506-0055}
\affiliation{Lawrence Berkeley National Laboratory, Berkeley, California 94720}
\affiliation{Lehigh University, Bethlehem, Pennsylvania 18015}
\affiliation{Max-Planck-Institut f\"ur Physik, Munich 80805, Germany}
\affiliation{Michigan State University, East Lansing, Michigan 48824}
\affiliation{National Research Nuclear University MEPhI, Moscow 115409}
\affiliation{National Institute of Science Education and Research, HBNI, Jatni 752050, India}
\affiliation{National Cheng Kung University, Tainan 70101 }
\affiliation{Nuclear Physics Institute of the CAS, Rez 250 68, Czech Republic}
\affiliation{Ohio State University, Columbus, Ohio 43210}
\affiliation{Institute of Nuclear Physics PAN, Cracow 31-342, Poland}
\affiliation{Panjab University, Chandigarh 160014, India}
\affiliation{Pennsylvania State University, University Park, Pennsylvania 16802}
\affiliation{NRC "Kurchatov Institute", Institute of High Energy Physics, Protvino 142281}
\affiliation{Purdue University, West Lafayette, Indiana 47907}
\affiliation{Rice University, Houston, Texas 77251}
\affiliation{Rutgers University, Piscataway, New Jersey 08854}
\affiliation{Universidade de S\~ao Paulo, S\~ao Paulo, Brazil 05314-970}
\affiliation{University of Science and Technology of China, Hefei, Anhui 230026}
\affiliation{Shandong University, Qingdao, Shandong 266237}
\affiliation{Shanghai Institute of Applied Physics, Chinese Academy of Sciences, Shanghai 201800}
\affiliation{Southern Connecticut State University, New Haven, Connecticut 06515}
\affiliation{State University of New York, Stony Brook, New York 11794}
\affiliation{Instituto de Alta Investigaci\'on, Universidad de Tarapac\'a, Arica 1000000, Chile}
\affiliation{Temple University, Philadelphia, Pennsylvania 19122}
\affiliation{Texas A\&M University, College Station, Texas 77843}
\affiliation{University of Texas, Austin, Texas 78712}
\affiliation{Tsinghua University, Beijing 100084}
\affiliation{University of Tsukuba, Tsukuba, Ibaraki 305-8571, Japan}
\affiliation{United States Naval Academy, Annapolis, Maryland 21402}
\affiliation{Valparaiso University, Valparaiso, Indiana 46383}
\affiliation{Variable Energy Cyclotron Centre, Kolkata 700064, India}
\affiliation{Warsaw University of Technology, Warsaw 00-661, Poland}
\affiliation{Wayne State University, Detroit, Michigan 48201}
\affiliation{Yale University, New Haven, Connecticut 06520}

\author{M.~S.~Abdallah}\affiliation{American University of Cairo, New Cairo 11835, New Cairo, Egypt}
\author{B.~E.~Aboona}\affiliation{Texas A\&M University, College Station, Texas 77843}
\author{J.~Adam}\affiliation{Brookhaven National Laboratory, Upton, New York 11973}
\author{L.~Adamczyk}\affiliation{AGH University of Science and Technology, FPACS, Cracow 30-059, Poland}
\author{J.~R.~Adams}\affiliation{Ohio State University, Columbus, Ohio 43210}
\author{J.~K.~Adkins}\affiliation{University of Kentucky, Lexington, Kentucky 40506-0055}
\author{G.~Agakishiev}\affiliation{Joint Institute for Nuclear Research, Dubna 141 980}
\author{I.~Aggarwal}\affiliation{Panjab University, Chandigarh 160014, India}
\author{M.~M.~Aggarwal}\affiliation{Panjab University, Chandigarh 160014, India}
\author{Z.~Ahammed}\affiliation{Variable Energy Cyclotron Centre, Kolkata 700064, India}
% ...

\collaboration{STAR Collaboration}\noaffiliation


\begin{abstract}
We report precision measurements of hypernuclei ${}^3_\Lambda \rm{H}$ and ${}^4_\Lambda \rm{H}$ lifetimes obtained from Au+Au collisions at \snn = 3.0\,GeV and 7.2\,GeV collected by the STAR experiment at RHIC, and the first measurement of ${}^3_\Lambda \rm{H}$ and ${}^4_\Lambda \rm{H}$ mid-rapidity yields in Au+Au collisions at \snn = 3.0\,GeV. ${}^3_\Lambda \rm{H}$ and ${}^4_\Lambda \rm{H}$, being the two simplest bound states composed of hyperons and nucleons, are cornerstones in the field of hypernuclear physics. Their lifetimes are measured to be $221\pm15(\rm stat.)\pm19(\rm syst.)$\,ps for ${}^3_\Lambda \rm{H}$ and $218\pm6(\rm stat.)\pm13(\rm syst.)$\,ps for ${}^4_\Lambda \rm{H}$. The $p_T$-integrated yields of ${}^3_\Lambda \rm{H}$ and ${}^4_\Lambda \rm{H}$ are presented in different centrality and rapidity intervals. It is observed that the shape of the rapidity distribution of ${}^4_\Lambda \rm{H}$ is different for 0--10\% and 10--50\% centrality collisions. Thermal model calculations, using the canonical ensemble for strangeness, describes the ${}^3_\Lambda \rm{H}$ yield well, while underestimating the ${}^4_\Lambda \rm{H}$ yield. Transport models, combining baryonic mean-field and coalescence (JAM) or utilizing dynamical cluster formation via baryonic interactions (PHQMD) for light nuclei and hypernuclei production, approximately describe the measured ${}^3_\Lambda \rm{H}$ and ${}^4_\Lambda \rm{H}$ yields. Our measurements provide means to precisely assess our understanding of the fundamental baryonic interactions with strange quarks, which can impact our understanding of more complicated systems involving hyperons, such as the interior of neutron stars or exotic hypernuclei. 
\end{abstract}

\maketitle
\realpagewiselinenumbers
\setlength\linenumbersep{0.10cm}

Hypernuclei are nuclei containing at least one hyperon. As such, they are excellent experimental probes to study the hyperon-nucleon 
($Y$--$N$) 
interaction. The $Y$--$N$ interaction is an important ingredient, not only in the equation-of-state (EoS) of astrophysical objects such as neutron stars, but also in the description of the hadronic phase of a heavy-ion collision~\cite{Steinheimer:2012tb}.
Heavy-ion collisions provide a unique laboratory to investigate the $Y$--$N$ interaction in finite temperature and density regions through the measurements of hypernuclei lifetimes, production yields etc. 

The lifetimes of hypernuclei ranging from $A=3$ to $56$ have previously been reported~\cite{Bhang:1998zp,PhysRev.136.B1803, Phillips:1969uy, PhysRev.139.B401, Bohm:1970xs,Acharya:2019qcp,Adam:2015yta,Rappold:2013fic,Abelev:2010rv, Adamczyk:2017buv}. 
The light hypernuclei ($A=3,4$), being simple hyperon-nucleon bound states, serve as cornerstones of our understanding of the $Y$--$N$ interaction~\cite{naturerecphys,Gal:2016boi}. For example, their binding energies $B_{\Lambda}$ are often utilized to deduce the strength of the $Y$--$N$ potential~\cite{Dalitz:1972vzj,PhysRev.137.B294,Kolesnikov:2006nu}, which is estimated to be roughly $2/3$ of the nucleon-nucleon potential. In particular, the hypertriton \hyt, a bound state of $\Lambda pn$, has a very small $B_{\Lambda}$ of several hundred keV~\cite{Juric:1973zq,STAR:2019wjm}, suggesting that the \hyt lifetime is close to the free-$\Lambda$ lifetime $\tau_{\Lambda}$. Recently, STAR~\cite{Abelev:2010rv, Adamczyk:2017buv}, ALICE~\cite{Acharya:2019qcp,Adam:2015yta} and HypHI~\cite{Rappold:2013fic} have reported \hyt lifetimes with large uncertainties ranging from $\sim50\%$ to $\sim100\%$ $\tau_{\Lambda}$. The tension between the measurements has led to debate~\cite{Perez-Obiol:2020qjy}. In addition, recent experimental observations of two-solar-mass neutron stars~\cite{Demorest:2010bx,Antoniadis:2013pzd,Barr:2016vxv} are incompatible with model calculations of the EoS of high baryon density matter, which predict hyperons to be a major ingredient in neutron star cores~\cite{Demorest:2010bx,Antoniadis:2013pzd,Barr:2016vxv}. These observations challenge our understanding of the $Y$--$N$ interaction, and call for more precise measurements~\cite{naturerecphys}. 

In heavy-ion collisions, particle production models such as statistical thermal hadronization~\cite{Andronic:2010qu} and coalescence~\cite{Steinheimer:2012tb} have been proposed to describe hypernuclei formation. While thermal model calculations primarily depend only on the freeze-out temperature and the baryo-chemical potential, the $Y$--$N$ interaction plays an important role in the coalescence approach, through its influence on the dynamics of hyperon transportation in nuclear medium~\cite{Feng:2020eqt}, as well as its connection to the coalescence criterion for hypernuclei formation from hyperons and nucleons~\cite{Steinheimer:2012tb}. At high collision energies, the \hyt yields have been measured by ALICE~\cite{Adam:2015yta} and STAR~\cite{Abelev:2010rv}. ALICE results from Pb+Pb collisions at \snn = 2.76\,TeV are consistent with statistical thermal model predictions~\cite{Andronic:2010qu} and coalescence calculations~\cite{Sun:2018mqq}. At low collision energies ($\sqrt{s_{\rm{NN}}}<20$ GeV), an enhancement in the hypernuclei yield is generally expected due to the higher baryon density~\cite{Steinheimer:2012tb,Andronic:2010qu}, although this has not been verified experimentally. The E864 and HypHI collaborations have reported hypernuclei cross sections at low collision energies~\cite{E864:2002xhb,Rappold:2015una}, however both measurements suffered from low statistics and lack of mid-rapidity coverage. Precise measurements of hypernuclei yields at low collision energies are thus critical to advance our understanding in their production mechanisms in heavy-ion collisions and to establish the role of hyperons and strangeness in the EoS in the high-baryon-density region~\cite{Chen:2018tnh}. In addition, such measurements provide guidance on searches for exotic strange matter such as double-$\Lambda$ hypernuclei and strange dibaryons in low energy heavy-ion experiments, which could lead to broad implications~\cite{CBM:2016kpk,jparc,Jaffe:1976yi}. 

In this letter, we report \hyt and \hyh lifetimes obtained from data samples of Au$+$Au collisions at \snn = 3.0\,GeV and 7.2\,GeV, as well as the first measurement of \hyt and \hyh differential yields at \snn = 3.0\,GeV. We focus on the yields at mid-rapidity in order to investigate hypernuclear production in the high-baryon-density region. The yields at \snn = 7.2\,GeV are not presented here due to the lack of mid-rapidity coverage.
The data were collected by the Solenoidal Tracker at RHIC (STAR)~\cite{STAR:2002eio} in 2018,
using the fixed-target (FXT) configuration. In the FXT configuration a single beam provided by RHIC impinges on a gold target of thickness 0.25\,mm (corresponding to a $1\%$ interaction probability) located at 201 cm away from the center of the STAR detector.
The minimum bias (MB) trigger condition is provided by the Beam-Beam Counters (BBC)~\cite{Whitten:2008zz} and the Time of Flight (TOF) detector~\cite{Llope:2012zz}. The reconstructed primary-vertex position along the beam direction is required to be within $\pm2$ cm of the nominal target position. The primary-vertex position in the radial plane is required to lie within a radius of 1.5\,cm from the center of the target to eliminate possible backgrounds arising from interactions with the vacuum pipe. 
In total, 2.8$\times 10^8$ (1.5$\times 10^8$) qualified events at \snn = 3.0 (7.2)\,GeV 
are used in this analysis. The \snn = 3.0\,GeV analysis and \snn = 7.2\,GeV analysis are similar. In the following, we describe the former; details related to the latter can be found in the supplementary material. 
 
The centrality of the collision is determined using the number of reconstructed charged tracks in the Time Projection Chamber (TPC)~\cite{Anderson:2003ur} compared to a Monte Carlo Glauber model simulation~\cite{Ray:2007av}. Details are given in~\cite{STAR:2020dav}. The top 0--50\% most central events are selected for our analysis. 
\hyt and \hyh are reconstructed via the two-body decay channels ${}^A_\Lambda \rm{H}\rightarrow \pi^{-} + {}^A \rm{He}$, where $A=3,4$. 
Charged tracks are reconstructed using the TPC in a 0.5\,Tesla uniform magnetic field. We require the reconstructed tracks to have at least 15 measured space points in the TPC (out of 45) and a minimum reconstructed transverse momentum of 150\,MeV$/c$ to ensure good track quality. Particle identification for $\pi^{-}$, $^{3}$He, and $^{4}$He is achieved by 
the measured ionization energy loss 
in the TPC. The KFParticle package~\cite{kfptc}, 
a particle reconstruction package based on the Kalman filter utilizing the error matrices,
is used for the reconstruction of the mother particle. 
Various topological variables such as the decay length of the mother particle, the distances of closest approach (DCA) between the mother/daughter particles to the primary vertex, and the DCA between the two daughters, are examined. Cuts on these topological variables are applied to the hypernuclei candidates in order to maximize the signal significance. In addition, we place fiducial cuts on the reconstructed particles to minimize edge effects. 

\begin{figure}[!htb]
\centering
\centerline{\includegraphics[width=0.5\textwidth]{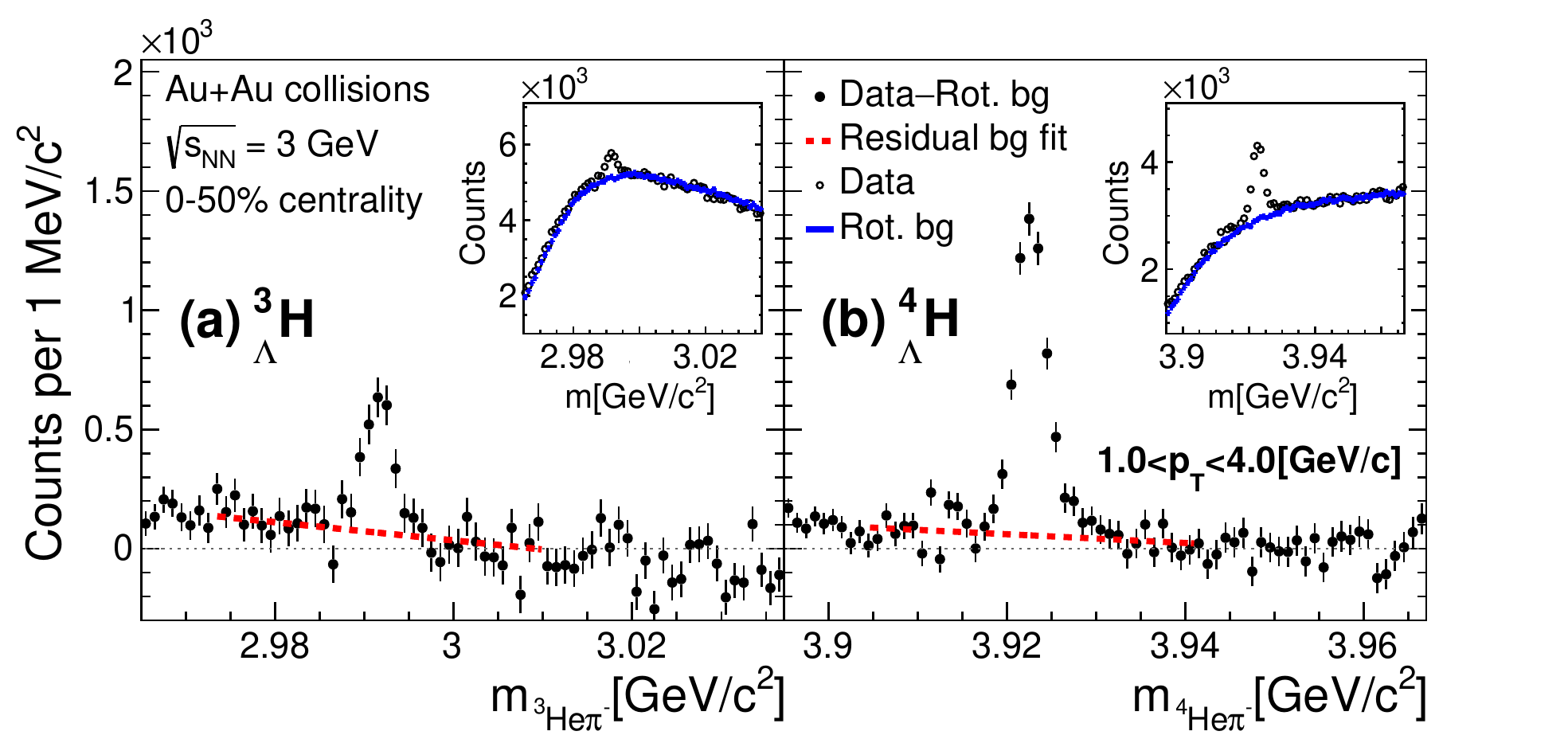}}
\centerline{\includegraphics[width=0.5\textwidth]{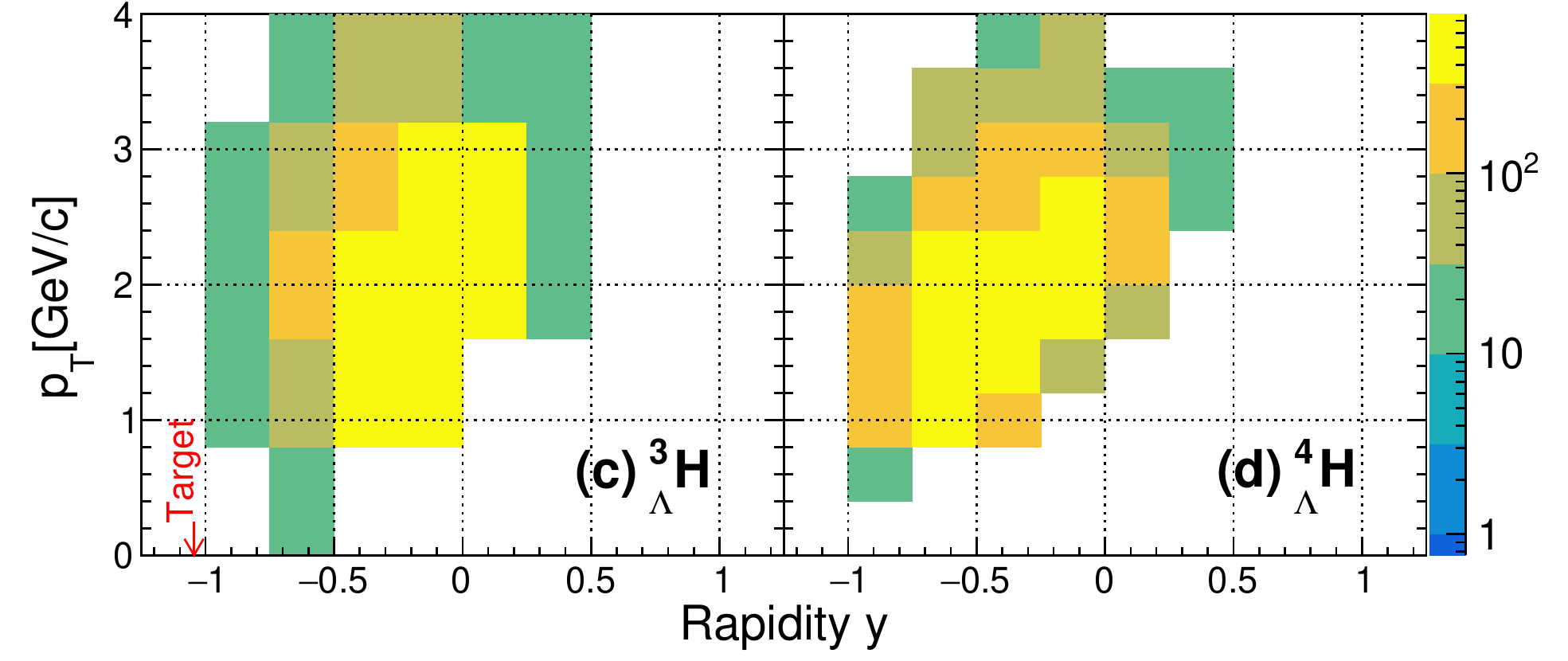}}

\caption{Top row: Invariant mass distributions of (a)$^{3}\rm{He}\pi^{-}$ and (b)$^{4}\rm{He}\pi^{-}$ pairs. In the insets, black open circles represent the data, blue histograms represent the background constructed by using rotated pion tracks. In the main panels, black solid circles represent the rotational background subtracted data, and the red dashed lines describe the residual background. Bottom row: The transverse momentum ($p_{T}$) versus the rapidity (y) for reconstructed (c)\hyt and (d)\hyh. The target is located at the $y = -1.05$.}
\label{fig1}
\end{figure}

Figure~\ref{fig1} (a,b) shows invariant mass distributions of ${}^{3}\rm{He}\pi^-$ pairs and ${}^{4}\rm{He}\pi^-$ pairs in the \pt region (1.0--4.0)\,GeV$/c$ for the 50\% most central collisions. The combinatorial background is estimated using a rotational technique, in which all $\pi^{-}$ tracks in a single event are rotated with a fixed angle multiple times and then normalized in the side-band region.  
The background shape is reasonably reproduced using this rotation technique for both \hyt and \hyh as shown in Fig.~\ref{fig1} (a,b). The combinatorial background is subtracted from the data in 2D phase space (\pt and rapidity $y$) in the collision center-of-mass frame. In addition to subtracting the rotational background, we perform a linear fit using the side-band region to remove any residual background. The subtracted distributions are shown in Fig.~\ref{fig1} (c,d). The target is located at $y$\,=\,$-1.05$, and the sign of the rapidity $y$ is chosen such that the beam travels in the positive $y$ direction. 
The mass resolution 
is 1.5 and 1.8 MeV$/c^{2}$ for \hyt and \hyh, respectively. 

The reconstructed \hyt and \hyh candidates are further divided into different $L/\beta\gamma$ intervals, where $L$ is the decay length, $\beta$ and $\gamma$ are particle velocity divided by the speed of light and Lorentz factor, respectively. 
The raw signal counts, $N^{\rm{raw}}$, for each $L/\beta\gamma$ interval are corrected
for the TPC acceptance, tracking, and particle identification efficiency, using an embedding technique in which the TPC response to Monte Carlo (MC) hypernuclei and their decay daughters is simulated in the STAR detector described in GEANT3~\cite{Brun:1987ma}. Simulated signals are embedded into the real data and processed through the same reconstruction algorithm as in real data.
The simulated hypernuclei, used for determining the efficiency correction, need to be re-weighted in 2D phase space $(p_T$--$y)$ such that the MC hypernuclei are distributed in a realistic manner. This can be constrained by comparing the reconstructed kinematic distributions ($p_{T}, y$) between simulation and real data. The corrected hypernuclei yield as a function of $L/\beta\gamma$ is fitted with an exponential function (see supplementary material) and
the decay lifetime is determined as the negative inverse of the slope divided by the speed of light. 

\begin{figure}[!htb]
\centering
\centerline{\includegraphics[width=0.5\textwidth]{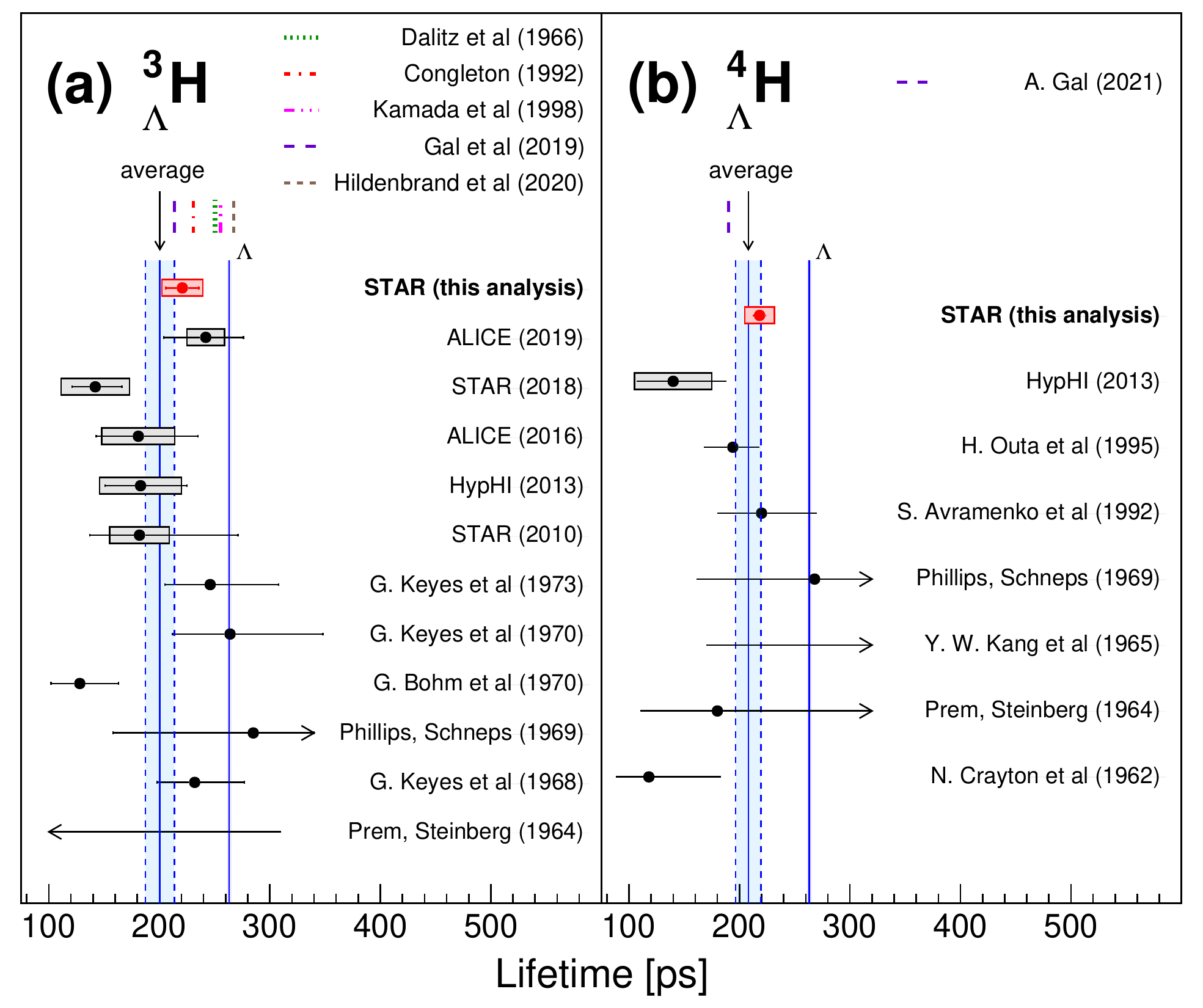}}
\caption{\hyt(a) and \hyh(b) measured lifetime, compared to previous measurements~\cite{Acharya:2019qcp,Adam:2015yta, Adamczyk:2017buv,Abelev:2010rv, Rappold:2013fic, Keyes:1974ev, Keyes:1970ck, Bohm:1970se, Phillips:1969uy, PhysRevLett.20.819,Avramenko:1992wy, PhysRev.136.B1803, Outa:1995es, PhysRev.139.B401, Crayton:1962oza}, theoretical calculations~\cite{Hildenbrand:2020kzu, Rayet:1966fe,Congleton:1992kk,Kamada:1997rv, Gal:2018bvq, galsqm} and $\tau_{\Lambda}$~\cite{Zyla:2020zbs}. Horizontal lines represent statistical uncertainties, while boxes represent systematic uncertainties. The experimental average lifetimes and the corresponding uncertainty of \hyt and \hyh are also shown as vertical blue shaded bands.}
\label{fig2}
\end{figure}





We consider four major sources of systematic uncertainties in the lifetime result: imperfect description of topological variables in the simulations, imperfect knowledge of the true kinematic distribution of the hypernuclei, the TPC tracking efficiency, and the signal extraction technique.
Their contributions are estimated by varying the topological cuts, the MC hypernuclei $p_T$--$y$ distributions, the TPC track quality selection cuts and the background subtraction method. 
The possible contamination of the signal due to multi-body decays of $A>3$ hypernuclei is
estimated using MC simulations and found to be negligible ($<0.1\%$) within our reconstructed hypernuclei mass window.
The systematic uncertainties
due to different sources are tabulated in Tab.~\ref{tab:urqmd_dv1}. They are assumed to be uncorrelated with each other and added in quadrature in the total systematic uncertainty. As a cross-check, we conducted the measurement of $\Lambda$ lifetime from the same data and the result is consistent with the PDG value~\cite{Zyla:2020zbs}(see supplementary material). 

\begin{table}[!htbp]
    \centering
    \begin{tabular}{c|cc|cc}
    \hline \hline
    & \multicolumn{2}{c}{Lifetime}  & \multicolumn{2}{|c}{$dN/dy$} \\
        \textbf{Source
        } & \textbf{$^{3}_{\Lambda}\rm{H}$} & \textbf{$^{4}_{\Lambda}\rm{H}$} 
        & \textbf{$^{3}_{\Lambda}\rm{H}$} & \textbf{$^{4}_{\Lambda}\rm{H}$}\\ \hline
        Analysis cuts & $5.5\%$ & $5.1\%$ 
        & $15.1\%$ & $6.9\%$\\ 
        Input MC  & $3.1\%$ & $1.8\%$ 
        & $8.8\%$ & $3.8\%$\\ 
        Tracking efficiency & $5.0\%$ & $2.4\%$
        & $14.1\%$ & $5.2\%$\\ 
        Signal extraction & $1.5\%$ & $0.7\%$ 
        & $14.3\%$ & $7.7\%$\\ 
        Extrapolation & N/A & N/A  
        & $13.6\%$ & $10.9\%$\\ 
        Detector material & $<1\%$ & $<1\%$ 
        & $4.0\%$ & $2.0\%$\\ \hline
        \textbf{Total} & \textbf{~~$8.2\%$~~} &  \textbf{~~$6.0\%$~~} &
        \textbf{~~$31.9\%$~~} & \textbf{~~$16.6\%$~~}\\ \hline \hline

    \end{tabular}
    \caption{Summary of systematic uncertainties for the lifetime and top 10\% most central $dN/dy$ ($|y|$$<$0.5) measurements using $\sqrt{s_{\rm{NN}}}$ = 3.0\,GeV data.}
    \label{tab:urqmd_dv1}
\end{table}

The lifetime results measured at \snn = 3.0\,GeV and \snn = 7.2\,GeV are found to agree well with each other. The combined results are
$221 \pm 15 (\rm stat.) \pm 19 (\rm syst.)$ ps for \hyt and 
$218 \pm  6 (\rm stat.) \pm 13 (\rm syst.)$ ps for \hyh. As shown in Fig.~\ref{fig2}, they are consistent with previous measurements from ALICE~\cite{Acharya:2019qcp,Adam:2015yta}, STAR~\cite{Adamczyk:2017buv,Abelev:2010rv}, HypHI~\cite{ Rappold:2013fic} and early experiments using imaging techniques~\cite{Keyes:1974ev, Keyes:1970ck, Bohm:1970se, Phillips:1969uy, PhysRevLett.20.819,Avramenko:1992wy, PhysRev.136.B1803, Outa:1995es, Abelev:2010rv, PhysRev.139.B401, Crayton:1962oza}. Using all the available experimental data, the average lifetimes of \hyt and \hyh are $200\pm13$ ps and $208\pm12$ ps, respectively, corresponding to $(76\pm5)\%$ and $(79\pm5)\%$ of $\tau_{\Lambda}$. All data from ALICE, STAR and HypHI lie within $1.5\sigma$ of the global averages. These precise data clearly indicate that the \hyt and \hyh lifetimes are considerably lower than $\tau_{\Lambda}$. 

Early theoretical calculations of the \hyt lifetime typically give values within $15\%$ of $\tau_{\Lambda}$~\cite{Rayet:1966fe,Congleton:1992kk,Kamada:1997rv}. This can be explained by the loose binding of $\Lambda$ in the \hyt. A recent calculation~\cite{Hildenbrand:2020kzu} using a pionless effective field theory approach with $\Lambda d$ degrees of freedom gives a \hyt lifetime of $\approx 98\%$ $\tau_{\Lambda}$. Meanwhile, it is shown in recent studies that incorporating attractive pion final state interactions, which has been previously disregarded, decreases the \hyt lifetime by $\sim15\%$~\cite{Gal:2018bvq,Perez-Obiol:2020qjy}. This leads to a prediction of the \hyt lifetime to be $(81\pm2)\%$ of $\tau_{\Lambda}$, consistent with the world average.

For \hyh, a recent estimation~\cite{galsqm} based on the empirical isospin rule~\cite{Cohen:1990qe} agrees with the data within $1\sigma$. The isospin rule is based on the  experimental ratio $\Gamma(\Lambda\rightarrow n+\pi^{0})/\Gamma(\Lambda\rightarrow p+\pi^{-}) \approx 0.5$, which leads to the prediction $\tau(^{4}_{\Lambda}\rm{H})/\tau(^{4}_{\Lambda}\rm{He})= (74\pm4)\%$~\cite{galsqm}. Combining the average value reported here and the previous $^{4}_{\Lambda}\rm{He}$ lifetime measurement~\cite{Outa:1998tt,Parker:2007yr}, the measured ratio $\tau(^{4}_{\Lambda}\rm{H})/\tau(^{4}_{\Lambda}\rm{He})$ is $(83\pm6)\%$, consistent with the expectation.

Previous measurements on light nuclei suggest that their production yields in heavy-ion collisions may be related to their internal nuclear structure~\cite{E864:1999zqh}. Similar relations for hypernuclei are suggested by theoretical models~\cite{Steinheimer:2012tb}. To further examine the hypernuclear structure and its production mechanism in heavy-ion collisions, we report the first measurement of hypernuclei $dN/dy$ in two centrality selections: top 0--10\% most central and 10--50\% mid-central collisions. The \pt spectra can be found in the supplementary material, and are extrapolated down to zero \pt to obtain the $p_{T}$-integrated $dN/dy$.
Different functions~\cite{STAR:2008med} are used to estimate the systematic uncertainties in the unmeasured region, which correspond to 32--60\% of the $p_T$-integrated yield in various rapidity intervals, and introduce 8--14\% systematic uncertainties. 
Systematic uncertainties associated with analysis cuts, tracking efficiency, and signal extraction
are estimated using
the same method as for the lifetime measurement.
We further consider the effect of the uncertainty in the simulated
hypernuclei lifetime on the calculated reconstruction efficiency by
varying the simulation's lifetime assumption within a 1$\sigma$ window of the average experimental lifetime, which leads to $8\%$ and $4\%$ uncertainty for \hyt and \hyh, respectively. Finally, 
hypernuclei may encounter Coulomb dissociation when traversing the gold target. The survival probability is estimated using a Monte Carlo method according to~\cite{Lyuboshitz:2007zzd}. The results show the survival probability $>96(99)\%$ for \hyt(\hyh) in the kinematic regions considered for the analysis. The dissociation has a strong dependence on $B_{\Lambda}$ of the hypernuclei.
Systematic uncertainties are estimated by varying the $B_{\Lambda}$ of the \hyt and \hyh, which are equal to $0.27\pm0.08$\,MeV and $2.53\pm 0.04$\,MeV, respectively~\cite{Liu:2019mlm}. As a conservative estimate, we assign the systematic uncertainty by comparing the calculation using the central values of $B_{\Lambda}$ and its 2.5$\sigma$ limits. A summary of the systematic uncertainties for the $dN/dy$ measurement is listed in Tab.~\ref{tab:urqmd_dv1}. 

\begin{figure}[!htb]
\centering
\centerline{\includegraphics[width=0.5\textwidth]{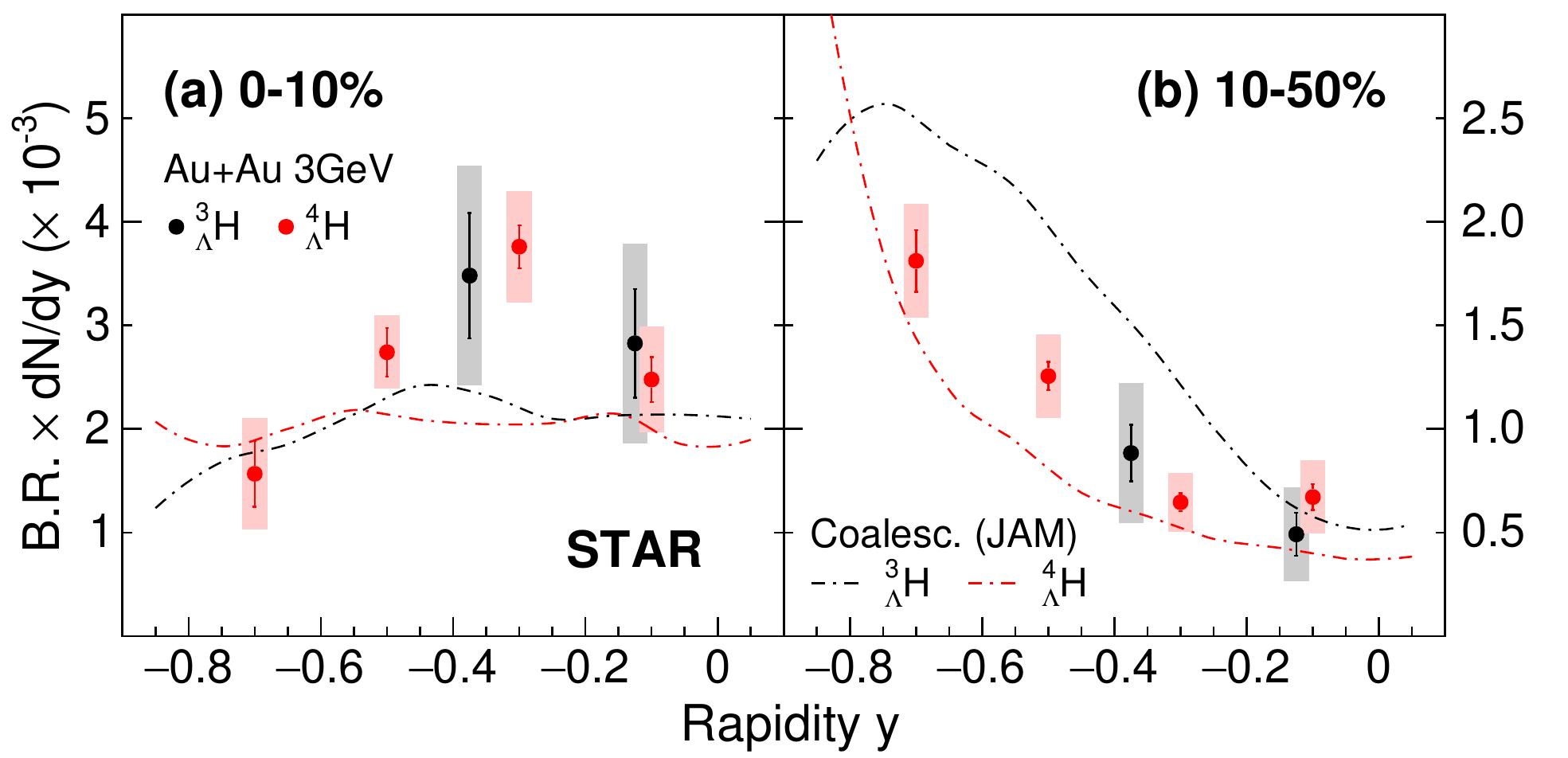}}
\caption{B.R.$\times dN/dy$ as a function of rapidity $y$ for \hyt(black circles) and \hyh(red circles) for (a) $0-10\%$ centrality and (b) $10-50\%$ centrality Au$+$Au collisions at $\sqrt{s_{\rm{NN}}} =$ 3.0 GeV. Vertical lines represent statistical uncertainties, while boxes represent systematic uncertainties. The dot-dashed lines represent coalescence (JAM) calculations. The coalescence parameters used are indicated in the text.}
\label{fig3}
\end{figure}

The \pt-integrated yields of \hyt and \hyh times the branching ratio (B.R.) as a function of $y$ are shown in Fig.~\ref{fig3}. For \hyh, we can see that the mid-rapidity distribution changes from convex to concave from 0--10\% to 10--50\% centrality. This change in shape is likely related to the change in the collision geometry, such as spectators playing a larger role in non-central collisions. 

Also shown in Fig.~\ref{fig3} are calculations from the transport model, JET AA Microscopic Transportation Model (JAM)~\cite{PhysRevC.61.024901} coupled with a coalescence prescription to all produced hadrons as an afterburner~\cite{Liu:2019nii}. In this model, deuterons and tritons are formed through the coalescence of nucleons, and subsequently, \hyt and \hyh are formed through the coalescence of $\Lambda$ baryons with deuterons or tritons. Coalescence takes place if the spatial coordinates and the relative momenta of the constituents are within a sphere of radius $(r_{C},p_{C})$. It is found that calculations using coalescence parameters $(r_{C},p_{C})$ of $(4.5\rm{fm},0.3\rm{GeV/}c)$, $(4\rm{fm},0.3\rm{GeV/}c)$, $(4\rm{fm},0.12\rm{GeV/}c)$ and $(4\rm{fm},0.3\rm{GeV/}c)$ for $d$, $t$, \hyt and \hyh respectively can qualitatively reproduce the centrality and rapidity dependence of the measured yields. The smaller $p_{C}$ parameter used for \hyt formation is motivated by its much smaller $B_\Lambda$ ($\sim0.3$MeV) compared to \hyh ($\sim2.6$MeV). The data offer first quantitative input on the coalescence parameters for hypernuclei formation in the high baryon density region, enabling more accurate estimations of the production yields of exotic strange objects, such as strange dibaryons~\cite{Steinheimer:2012tb}. 

\begin{figure}[!htb]
\centering
\centerline{\includegraphics[width=0.42\textwidth]{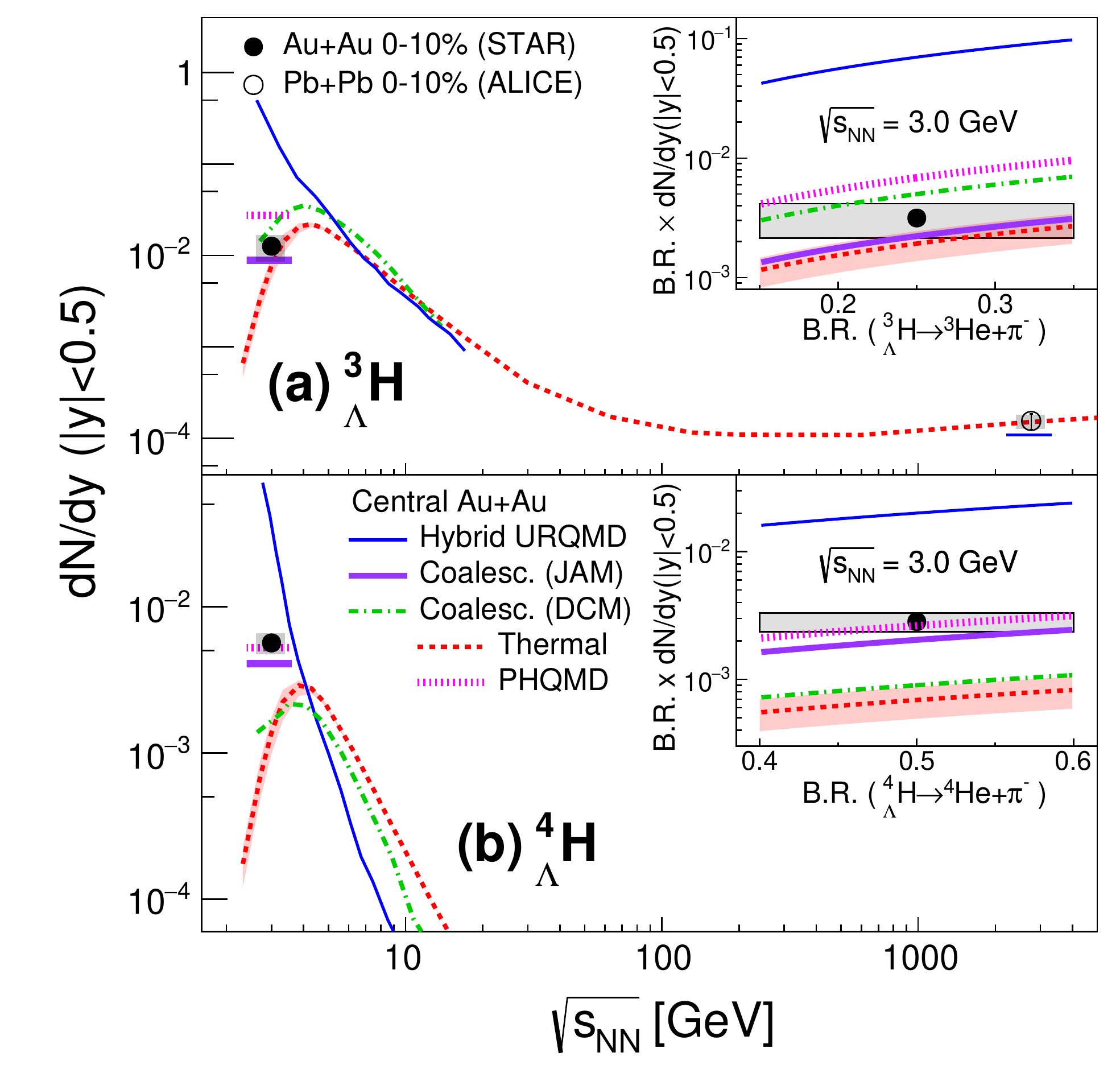}}
\caption{(a) \hyt and (b) \hyh yields at $|y|<0.5$ as a function of beam energy in central heavy-ion collisions. The symbols represent measurements~\cite{Adam:2015yta} while the lines represent different theoretical calculations. The data points assume a B.R. of $25(50)$\% for $^{3}_{\Lambda}\rm{H}(^{4}_{\Lambda}\rm{H})\rightarrow {}^{3}\rm{He}(^{4}\rm{He})+\pi^{-}$. The insets show the (a) \hyt and (b) \hyh yields at $|y|<0.5$ times the B.R. as a function of the B.R.. Vertical lines represent statistical uncertainties, while boxes represent systematic uncertainties.}
\label{fig4}
\end{figure}

The decay B.R. of $^{3}_{\Lambda}\rm{H}\rightarrow\rm{^{3}He} + \pi^{-}$ was not directly measured. A variation in the range $15-35\%$ for the B.R.~\cite{Congleton:1992kk,Adamczyk:2017buv,Kamada:1997rv} is considered when calculating the total $dN/dy$. For $^{4}_{\Lambda}\rm{H}\rightarrow\rm{^{4}He} + \pi^{-}$, a variation of $40-60\%$ based on~\cite{Outa:1998tt,Juric:1973zq} is considered in this analysis.

The \hyt and \hyh mid-rapidity yields for central collisions as a function of center-of-mass energy are shown in Fig.~\ref{fig4}. The uncertainties on the B.R.s are not shown in the main panels. Instead, the insets show the $dN/dy \times$B.R. as a function of B.R.. We observe that the \hyt yield at \snn = 3.0\,GeV is significantly enhanced compared to the yield at \snn = 2.76\,TeV~\cite{Adam:2015yta}, likely driven by the increase in baryon density at low energies. 

Calculations from the thermal model, which adopts the canonical ensemble for strangeness~\cite{Andronic:2010quprelim} that is mandatory at low beam energies~\cite{STAR:2021hyx} are compared to data. Uncertainties arising from the strangeness canonical volume are indicated by the shaded red bands. $\gamma$-decay of the excited state ${}^{4}_{\Lambda}\rm{H}(1^{+})$ to the ground state is accounted for in this calculation. Interestingly, while the \hyt yields at \snn = 3.0\,GeV and 2.76\,TeV are well described by the model, the \hyh yield is underestimated by approximately a factor of 4. Coalescence calculations using DCM, an intra-nuclear cascade model to describe the dynamical stage of the reaction~\cite{Steinheimer:2012tb}, are consistent with the \hyt yield while underestimating the \hyh yield, whereas the coalescence (JAM) calculations are consistent with both. We note that in the DCM model, the same coalescence parameters are assumed for \hyt and \hyh, while in the JAM model, parameters are tuned separately for \hyt and \hyh to fit the data. It is expected that the calculated hypernuclei yields depend on the choice of the coalescence parameters~\cite{Steinheimer:2012tb}. Recent calculations from PHQMD~\cite{Glassel:2021rod,Kireyeu:2019szf}, a microscopic transport model which utilizes a dynamical description of hypernuclei formation, is consistent with the measured yields within uncertainties. Compared to the JAM model which adopts a baryonic mean-field approach, baryonic interactions in PHQMD are modelled by density dependent 2-body baryonic potentials. Meanwhile, the UrQMD-hydro hybrid model overestimates the yields at \snn = 3.0\,GeV by an order of magnitude. Our measurements possess distinguishing power between different production models, and provide new baselines for the strangeness canonical volume in thermal models and coalescence parameters in transport-coalescence models. Such constraints can be utilized to improve model estimations on the production of exotic strange matter in the high baryon density region. 

In summary, precise measurements of \hyt and \hyh lifetimes have been obtained using the data samples of Au$+$Au collisions at \snn = 3.0 and 7.2\,GeV. The lifetimes are measured to be $221 \pm 15 (\rm stat.) \pm 19 (\rm syst.)$ ps for \hyt and $218 \pm 6 (\rm stat.) \pm 13 (\rm syst.)$ ps for \hyh. The averaged \hyt and \hyh lifetimes combining all existing measurements are both smaller than $\tau_{\Lambda}$ by $\sim 20\%$. The precise \hyt lifetime reported here resolves the tension between STAR and ALICE. We also present the first measurement of rapidity density of \hyt and \hyh in 0--10\% and 10--50\% $\sqrt{s_{\rm{NN}}}=3.0$\,GeV Au+Au collisions. Hadronic transport models JAM and PHQMD calculations reproduce the measured midrapidity \hyt and \hyh yields reasonably well. Thermal model predictions are consistent with the \hyt yield. Meanwhile, the same model underestimates the \hyh yield. We observe that the \hyt yield at this energy is significantly higher compared to those at \snn = 2.76\,TeV. This observation establishes low energy collision experiments as a promising tool to study exotic strange matter.

\parskip 4mm
We thank the RHIC Operations Group and RCF at BNL, the NERSC Center at LBNL, and the Open Science Grid consortium for providing resources and support.  This work was supported in part by the Office of Nuclear Physics within the U.S. DOE Office of Science, the U.S. National Science Foundation, National Natural Science Foundation of China, Chinese Academy of Science, the Ministry of Science and Technology of China and the Chinese Ministry of Education, the Higher Education Sprout Project by Ministry of Education at NCKU, the National Research Foundation of Korea, Czech Science Foundation and Ministry of Education, Youth and Sports of the Czech Republic, Hungarian National Research, Development and Innovation Office, New National Excellency Programme of the Hungarian Ministry of Human Capacities, Department of Atomic Energy and Department of Science and Technology of the Government of India, the National Science Centre of Poland, the Ministry  of Science, Education and Sports of the Republic of Croatia and German Bundesministerium f\"ur Bildung, Wissenschaft, Forschung and Technologie (BMBF), Helmholtz Association, Ministry of Education, Culture, Sports, Science, and Technology (MEXT) and Japan Society for the Promotion of Science (JSPS).

\bibliography{reference.bib}
\newpage



\end{document}